# Robust Low-Bias Negative Differential Resistance in Graphene Superlattices

S. M. Sattari-Esfahlan, J. Fouladi-Oskuei and S. Shojaei [*]

*Photonics Department, Research Institute for Applied Physics and Astronomy (RIAPA),
University of Tabriz, 51665-163 Tabriz, Iran*

**Abstract**

In this work, we present a detailed theoretical study on low bias current-voltage (I-V) characteristic of biased planar graphene superlattice (PGSL) provided by heterostructured substrate and series of grounded metallic planes placed over graphene sheet which induce periodically modulated Dirac gap and Fermi velocity barrier, respectively. We investigate the effect of PGSL parameters on I-V characteristic and appearance of multipeak negative differential resistance (NDR) in proposed device within the Landauer-Buttiker formalism and adopted transfer matrix method. Moreover, we propose a novel venue to control the NDR in PGSL with Fermi velocity barrier. Different regimes of NDR have been recognized based on PGSL parameters and external bias. From this viewpoint, we obtain multipeak NDR through miniband aligning in PGSL. Maximum pick to valley ratio (PVR) up to 167 obtained for $v_c$, Fermi velocity correlation (ratio of Fermi velocity in barrier and well region), of 1.9 at bias voltages between 70-130 mV. While our findings have good agreement with those of experiments can be considered in multi-valued memory, functional circuit, low power and high-speed nanoelectronic devices application.

**Keywords:** Negative differential resistance (NDR); Planer graphene superlattices (PGSLs); Fermi velocity barrier; Dirac gap

*Corresponding author.
  E. mail addresses: s_shojaei@tabrizu.ac.ir, shojaei.sh@gmail.com (S. Shojaei).
Tel: +98 4133392995, Fax: +98 4133347050, Skype: Saeed.shojaei

## I. Introduction

Graphene, an atomically thin two-dimensional material, has attracted a great deal of interest due to its superior electrical and optical properties [1-3] which is become one of the most promising materials for nanoelectronic and nanophotonic devices [4-6]. The charge carriers in monolayer graphene behave like a massless Dirac fermions with linear energy dispersion, can be governed by an effective Dirac equation, which leads to many novel electronic and transport properties [7-10]. However, due to the lack of energy gap in the pristine graphene electronic band structure, the Klein tunneling prevents Dirac electrons in graphene from being confined by an electrostatic potential, which impedes its use in electronic devices. In order to overcome this limitation, several schemes have been proposed which leads to suppression of Klein tunneling and confinement of the Dirac electrons in graphene. For instance, a energy gap can be induced by doping [7-9], substrate [10-12], strain [13-15], quantum confinement effects [16-19], spatial modulation of Dirac gap and external periodic potentials [20-23].The Fermi velocity is another key concept to study electronic properties of graphene. According to Lorentz invariant theory, by controlling the electron-electron interaction, one can regulate Fermi velocity in graphene. This implements various procedures such as dielectric screening [24, 25]; carrier concentration [26-28]; periodic potentials [29]; curvature of the graphene sheet [30] and substrate modulation [31]. Recently, graphene superlattice, graphene under periodic potentials, has been extensively studied both experimentally and theoretically [20-23, 32-38]. These works have considered periodic potentials of a different nature (electric & magnetic) and different profiles. This considerable interest is motivated by discovering novel transport properties of charge carriers through PGSL structures (e.g. extra Dirac points, highly anisotropic Dirac points) which has not been observed in the pristine graphene, and also by promising potential applications in a diverse area of nanoelectronic devices.

Beyond the usual linear or saturation performances expected to occur in graphene based devices, an appearance of prominent non-linear effects such as the NDR in the current–voltage characteristics has attracted strong research interest both theoretically and experimentally [39-51], since it could potentially impact the number of key applications such as high frequency oscillators, reflection amplifiers, memories, multi-level logic devices, and fast switches [52].

NDR at high bias voltages (1–2 V), in narrow nanoribbons have been studied in lots of works [46, 47]. NDR at low bias regime can also be achieved in monolayer and bilayer graphene, and graphene nanoribbon superlattices systems [48-50].

In present study, for a first time, we investigate the NDR features of graphene based planar superlattices composed of periodically modulated Fermi velocity, Dirac gap and electrostatic potential at low bias voltages $V_{SD}$<200 mV , above mentioned effects are induced by means of velocity barrier, nanostructured substrate and one-dimensional potentials of square barriers, respectively. Transmission properties of Dirac fermion beams tunneling through PGSL are investigated by adopted transfer matrix method (TMM) in details. We have found that the transmission properties of PGSL can be tuned readily by changing the main parameters of the PGSL, i.e. well and barrier widths, energy and angle of the incident electrons, the number of periods of PGSLs, Dirac gap and Fermi velocity barrier. Current-voltage characteristics obtained within the Landauer-Buttiker formalism. We have found that the NDR appears with appropriate structural parameters, which means that NDR could be controlled and tuned by PGSL parameters. More interestingly, we investigated the effect of Fermi velocity barrier on I-V characteristics of PGSL resonant tunneling diode (PGSLRTD).We found that PVR in PGSLRTD is enhanced exponentially by increasing Fermi velocity correlation.

## II. Model & Method

A schematic view of our proposed device is shown in Fig. 1. A monolayer of graphene is placed on planar heterostructured substrate composed of two different materials shown by I,II. This composed substrate, opens different energy gaps in different regions of the graphene sheet, denoted by $\Delta_W$ and $\Delta_B$. W, B refer to well and barrier, respectively. Moreover, according to Lorentz invariant theory, Fermi velocity differs in well & barrier regions of graphene placed on different materials of substrate [52]; thus, composed substrates can also modifies Fermi velocity in graphene. The Fermi velocity in each region will be denoted by $v_W$ and $v_B$. We define $v_c = v_B/v_W$ is the Fermi velocity correlation between barrier and well region. Series of grounded metallic planes are placed over graphene sheet which induce a periodic velocity barrier. Metallic planes do electron-electron interaction weaker; therefore reduce Fermi velocity in the relevant region [25]. Metal electrodes in left and right create electrostatic potential across the graphene sheet acting as source-drain voltage, $V_b = V_R - V_L$.

The Kronig-Penney (KP) model is applied for investigation of PGSL electrostatic potential profile. In the vicinity of the Dirac points, the PGSL electronic structure can be described by the Dirac-like equation. The effective 2D Dirac Hamiltonian for a PGSL with a position dependent energy gap and Fermi velocity, $v_F$, is written as:

$$H = -i\hbar\left(\sqrt{v_F(x)}\sigma_x \partial_x \sqrt{v_F(x)} + v_F(x)\sigma_y \partial_y\right) + V(x)\hat{1} + \Delta(x)\sigma_z \qquad (1)$$

Where $\sigma_i$ are the Pauli matrices and $\hat{1}$ is the 2× 2 unitary matrix, $\Delta(x)$ is the position dependent graphene energy gap, $V(x)$ is an external position dependent electrostatic potential that is composed of two parts: first, KP potential ($V_{KP}$) that takes as zero in well region and 400meV in barrier region. Second part indicates the applied external potential bias that is taken as *eEx* (*e* is electron charge, *E* is electric field and *x* is growth direction of PGSL)

$$V(x) = \begin{cases} V_B(x) = V_{KP}^B - eEx & \text{for barrier} \\ V_W(x) = V_{KP}^W - eEx & \text{for well,} \end{cases} \qquad (2)$$

Where indices W, B refer to well, barrier regions. According to Bloch's theorem, it is straightforward to obtain the electronic dispersion for periodic structure of PGSL by following equation as:

$$cos(k_x l) = cos(k_B d_B) cos(k_W d_W) + \frac{k_y^2 \hbar^2 v_B v_W - (E-V_B(x))(E-V_W(x)) + \Delta_B \Delta_W}{\hbar^2 v_B v_W k_B k_W} sin(k_B d_B) sin(k_W d_W) \quad (3)$$

Where $k_W = (((E-V_W(x))^2 - \Delta_W^2)/\hbar^2 v_W^2 - k_y^2)^{1/2}$, $k_B = (((E-V_B(x))^2 - \Delta_B^2)/\hbar^2 v_B^2 - k_y^2)^{1/2}$, $k_y = (Esin\theta/\hbar v_F)^{1/2}$, $\theta$ is incident angle of electron beam that is defined as angle between growth direction of planar PGSL and direction of incidence, $d_w$ and $d_B$ are the well and barrier width, respectively. We choose $V_{KP}^W = 0$, $V_{KP}^B = 400$ meV, $\Delta_B \neq 0$, $\Delta_w = 0$.

In order to adopt the well known TMM to the graphene superlattice under bias in this work, we consider m layers containing barrier and well with the potential mentioned in eq. 2 that can be rewritten as:

$$V_m = V_m^{KP} - eV_{RL}x_m/L. \quad (4)$$

Index m refers to number of layers (m=1,2,3,…N). $V_{RL}$ is difference of potential applied between two left and right end electrodes.

The solution to Dirac equation, $H\psi(x,y) = E\psi(x,y)$ is two component spinor that illustrates the two graphene sublattices reads:

$$\psi_m(x,y) = \psi_m(x)e^{ik_y y}. \quad (5)$$

and defining $\sqrt{v_F(x)_m}\psi_m(x) = \phi_m(x)$, the solution to Dirac equation in x direction can be written as a linear combination of forward/backward plane-waves [55]:

$$\phi_m^1(x,y) = \sqrt{v_{F_m}(x)}(A_m e^{ik_m x} + B_m e^{-ik_m x})e^{ik_y y},$$

$$\phi_m^2(x,y) = \sqrt{v_{F_m}(x)}(A_m e^{ik_m x + i\alpha_m} + B_m e^{-ik_m x - i\alpha_m})e^{ik_y y}, \quad (6)$$

where $A_m$ and $B_m$ are the transmission amplitudes, and $k_y = (Esin\theta/\hbar v_F)^{1/2}$. $\theta$ is incident angle of electron beam that is defined as angle between growth direction of planar PGSL and direction of incidence. $\alpha_m = tan^{-1}(k_y/k_m)$. The external potential in our work is linear with x. To construct the transfer matrix for this structure we divided the linear potential to very small potential steps where transfer matrix of each small step in mth region can be written simply through continuity conditions between spinors as [56]:

$$M_j = \begin{pmatrix} e^{ik_m^j x} & e^{-ik_m^j x} \\ e^{ik_m^j x + i\alpha_m} & -e^{-ik_m^j x - i\alpha_m} \end{pmatrix} \quad (7)$$

$$M_m(x) = \prod_j M_j(x)$$

Where, index j refers to jth division in mth region of PGSL. The transfer matrix of whole system, connects left electrode to right electrode, is obtained as:

$$t = \prod_m M_m^{-1}(x_{m,m+1}) M_{m+1}(x_{m,m+1}) \quad (8)$$

From the current density in mth region, $J_m(x) = v_{F_m}(x)\varphi_m^\dagger(x)\sigma_x\varphi_m(x)$, The current flow must be conservative at left and right sides of PGSL that implies $J_L(x) = J_R(x)$. From which we identify the transmission coefficient, $T_{LR}$[43, 53].:

$$T_{LR}(E, k, V_b) = \left|\frac{J_R}{J_L}\right| = \frac{\cos\theta_R}{\cos\theta_L}|t|^2. \tag{9}$$

In our model, the potential drop from source to drain follows a linear function. The zero-temperature ballistic net current as a function of bias is computed by the Landauer-Büttiker formalism [57- 58]:

$$I = I_0 \int_{\mu_R}^{\mu_L} dE |E| \times \int_{-\pi/2}^{\pi/2} \mathcal{T}(E_F, \alpha, v_b) \cos\alpha \, d\alpha \tag{10}$$

Where $I_0 = 2geW/v_F h^2$, $g = 4$ is the degeneracy of electron states in graphene, $W$ is the sample width that we choose it as $d_B$ and $V_b$ is applied voltage between the left and right electrode ($V_{bias}$ in Fig. 1), a linear voltage along the x-direction defined as $eV_b = \mu_L - \mu_R$ and $\mu_L(\mu_R)$ is the bias-dependent local Fermi energy in the left (right) electrode.

### III. Results & Discussions

Due to significant roles of device dimension and geometry in I-V characteristic, first, we analyze the effect of well and barrier width, $d_W, d_B$, and number of periodicity, N, on NDR. In Fig. 2 we present I-V diagram of device for different values of $d_W$ and certain value of $d_B$. We consider well width bigger than barrier width, and N=10.

As it can be seen from Fig. 2, for wider wells, correspondent current peaks values are smaller than that of thinner wells. Also, peaks are shifted to lower biases. This behavior can be attributed by red shifting of resonant energies with increasing well width. In the case of wider wells, lower biases are needed to align the resonant states. We found that transmission oscillations increases with increasing well widths (not shown here) that is because of increasing resonant states. On the other hand, increasing oscillations reduces the area of under transmission curve that subsequently reduces the current value for wider wells(Eq. 4).We found that for barrier width remarkably bigger than and comparable with well width, applying bias aligns resonant modes, so that resonant sequential resonant tunneling occurs and current increases to reach a peak. By increasing bias more, the resonant states are misaligned and strong tunneling suppression appears leads to in current decreasing and pronounced spike in I-V curve.

Now, we turn our attention to the impact of N on NDR and discuss on its different distinct regimes in our system. Fig. 3.adepicts the transmission spectrum(t) of PGSL as function of N. From figure, it is observed that by increasing N and subsequently increasing interfaces, Fabry-Perot like interferences are increased and lots of resonant modes are created. Furthermore, resonant modes show up below and above the stop band for all values of N. In the case of thin barriers in Fig. 3.b, for small value of N as 2 current increases monotonically upon resonant modes are aligned and sequential resonant tunneling occurs. Higher external biases misalign theresonant modes and tunneling is completely suppressed results in decreasing current value and NDR appears as a peak (curve of N=10). This trend indicates classical regime.

For large number of N (10 and more), resonant modes form superlattice minibands. At low biases the current is dominated by transmission across these minibands, called miniband regime. With increasing bias, resonant modes, (responsible for formation of minibands),are misaligned that breaks up the minibands into off resonant Wannier Stark ladders with suppressed transmission leading to current decreasing. By increasing bias, rungs of ladders from distinct minibands cross showing new resonant peaks in transmission and system encounter Wannier-Stark regimes. Such alternation process from miniband to Wannier-Stark regime is the reason of multi-peak NDR.In the other words, this trend of NDR for larger number of barriers originates from destructive-constructive interferences of Dirac fermions creating several peaks in transmission spectrum and I-V diagram. Hereafter, we consider the case of miniband and Wannier-Stark regime by taking number of barriers as N=10.

It is worth to mention that Fabry-Pérot like resonances appear for non-perpendicular incident (i.e. $k_y \neq 0$ ) while for $k_y = 0$ Klein tunneling suppresses resonant tunneling and NDR. Fig. 4 illustrates Dirac barrier gap ($\Delta_B$) effect on I-V characteristic in low bias regime. Based on our calculations, with increasing Dirac barrier gap, wider stop bands in transmission spectrum causes decreasing of current. On the other hand, increasing Dirac barrier gap inhibits Dirac carrier transport meaning that PVR increases. Also, multi peak trend appears for bigger Dirac barrier gap caused by resonant tunneling enhancement. This finding has a good agreement with that of Song et al and Sollner et al [59, 60]. Appropriate structural parameters to reach maximum NDR obtained as $v_c = 1.9$, $V_b$=70-130mV, *N*=10, $d_W$=40nm and $d_B$=5nm.Considerable value of PVR as 167 obtained for Dirac barrier gap of 150meV that is very good value for optimal performance of RTD based on PGSL that can be realized experimentally. In order to compare with other works done on graphene based NDR, we consider reference 59 where Song et al reported maximum value of PVR as almost 2.They used double barrier RTD based on graphene strips to obtain PVR. The advantage of our model is obtaining very big PVR in ultra-low bias regime by the means of PGSL. In our work variety of parameters have been taken into account to control NDR and obtain high value of PVR that NDR appears for suitable values of them. For example, very high value of barrier width prohibits the tunneling of electrons and current increases with increasing bias voltage linearly leading to prohibition of NDR appearance. Optimization on parameters should be done to find high value of PVR

One of key parameters in I-V characteristic is the ratio of Fermi velocity in barrier/well regions ($v_c$). It is very important to study the effect of this parameter to control NDR. In Fig.5(a) we report the transmission coefficient spectrum for different values of $v_c$. It can be seen that transmission is enhanced for special value of $v_c$ and energies. Closer view of Fig 5.a determines that oscillatory behavior in transmission spectrum becomes more pronounced at specific values of $v_c$. In contrast, at higher values of $v_c$ oscillation diminishes and energy gap is increased. Fig. 5(b) depicts that big values of current are obtained for smaller $v_c$. This finding can be explained by, transmission spectrum in Fig. 5(a) where larger area under transmission curve can be observed that leads to bigger current (Eq.4). Two main peaks are observed in low bias regime around 70mV and 130mVthat implies to NDR appearance. Such phenomenon originate from this fact that bigger $v_C$ creates higher Fermi velocity barrier so that it is difficult for Dirac electrons to be able to tunnel from them leading to reduction of current. NDR is resulted by resonant tunneling occurring in miniband and Wannier-Stark regime as discussed before. To get closer insight, we focus on the trends of PVRs related to two main peaks versus $v_C, dw, N, \Delta_B$ in Fig.6. It is clear that PVR's values are increased with increasing these parameters except than N. Fig. 6.c illustrates that increasing number of periods, N, increases PVR1 but decreases PVR2. Optimum values of structural parameters to enhance the PVR at low biases can be obtained from Fig.6. Our results show that among structural parameters $v_C$, has remarkable effects on the value of PVR1 and PVR2, hence it is very important in characterization of electronic device through cut-off frequency which is fundamental parameter in high frequency PGSLRTD based oscillators and switches. Our obtained high value of PVR at low biases is desired for implementation innovel low power digital logic circuits and multi-valued memory.

Finally, in order to verify our results, we compared our findings with those of Antonova et. al[61].They investigated NDR in the films fabricated from partially fluorinated graphene suspension. The formation of Graphene islands (quantum dots) are observed in these films. Various types of negative differential resistance (NDR) and a step-like increase in the current are found for films created from the fluorinated graphene suspension. NDR resulted from the formation of the potential barrier system in the film and graphene quantum dots, was reported in their study. In our study, we model this system by quantum well/barrier systems according to our theoretical framework. I-V characteristic found by our calculation has very good agreement with I-V characteristic measured by them in low biases.

## IV. Conclusion

In summary, we theoretically studied the I-V characteristics of planar graphene superlattice modeled by patterning graphene on nanostructured substrate and top grounded metal planes. To reach this goal, we adopted transfer matrix method to obtain transmission spectrum under external bias. The effects of parameters such as number of periods, *N*, well and barrier width, $d_W, d_B$ and ratio of Fermi velocity in each region, $v_C$, Dirac barrier and electrostatic barrier have been analyzed in details to enhance peak to valley ratio, PVR, in I-V curve at very low external biases through within the Landauer-Buttiker formalism. Different regimes of NDR were discussed according to parameters used in our model device. Appropriate structural parameters to reach maximum NDR obtained as $v_C$ =1.9, $V_b$=70-130 mV, *N*=10, $d_w$= and $d_b$=5 nm. Robust value of PVR as 167 obtained for Dirac barrier gap of150 meVat ultra-low biases. Our proposed device paves a novel way in the PGSL based NDR devices such as RTDs, multilevel memory, low power and high-speed electronic devices working at low bias.

**Refrences**

**Figure caption**

Fig.1. (Color online) (a) Schematic view of Dirac-cone distribution and band-edge profile as Kronig-Penney model in typical PGSL in the presence of electric field. $\Delta_g$ shows gap opened in barrier/well regions, shown in our calculation as $\Delta_w$, $\Delta_b$ ,imposed by periodic substrate(b) Schematic diagram of the device containing top metal plates ( shown with yellow color) that create Fermi velocity barrier. Left drain and source electrodes labeled with left and right electrode respectively. $L$ and $W$ are length and width of device. $d_W, d_B$ are well, barrier width ,respectively.

Fig. 2. (Color online) I-V characteristic for three values of well width, $d_W$, for $d_B = 5\ nm$, N=10 and $\Delta_B = 150\ meV$.

Fig. 3 (Color online) (a) Map of Transmission spectrum for different values of N. Incident angle, $\theta$, is $pi/6$. (b) I-V characteristics for different values of N. $d_W = 40 nm$ , $\Delta_B = 150\ meV$ and $d_B = 5\ nm$.

Fig. 4. (Color online) The effect of gap opening in barrier region, $\Delta_B$, on I-V characteristic. N=10, $d_W = 40 nm, d_B = 5\ nm$.

Fig. 5. (Color online) (a) Counter plot of transmission spectrum for different values of correlation velocity, $v_C$ for incident angle of $\theta = pi/6$. (b) I-V characteristics for different correlation velocity. $\Delta_B = 150\ meV$, N=10, $d_W = 40\ nm$ and $d_B = 5\ nm$.

Fig. 6: (Color online) Variation of PVR1 and PVR2 for various structural parameters of PGSL (a)$v_c$, (b) $d_W$, (c)N and (d)$\Delta_B$. Constant parameters in each plot are considered as $\Delta_B = 150\ meV$, N=10, $d_W = 40\ nm$ and $d_B = 5\ nm$.

Fig1:

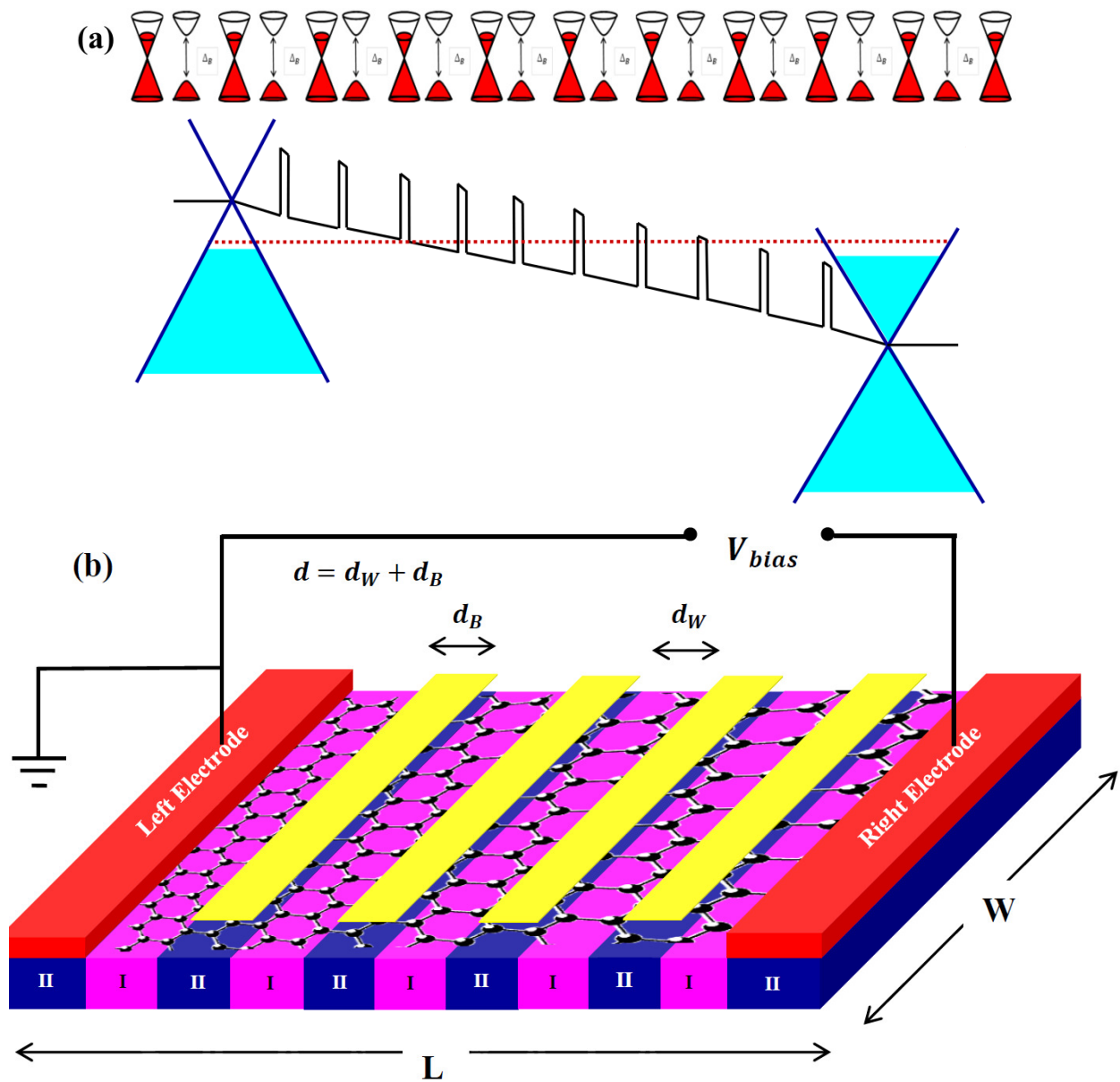

Fig. 2:

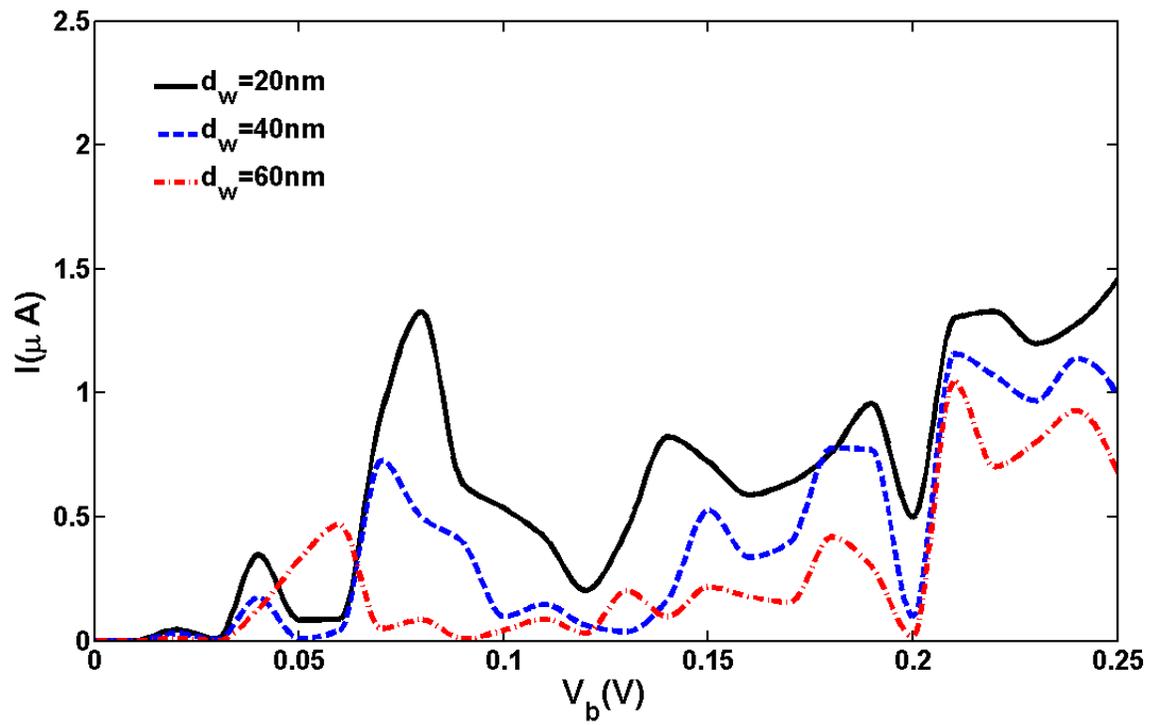

Fig. 3:

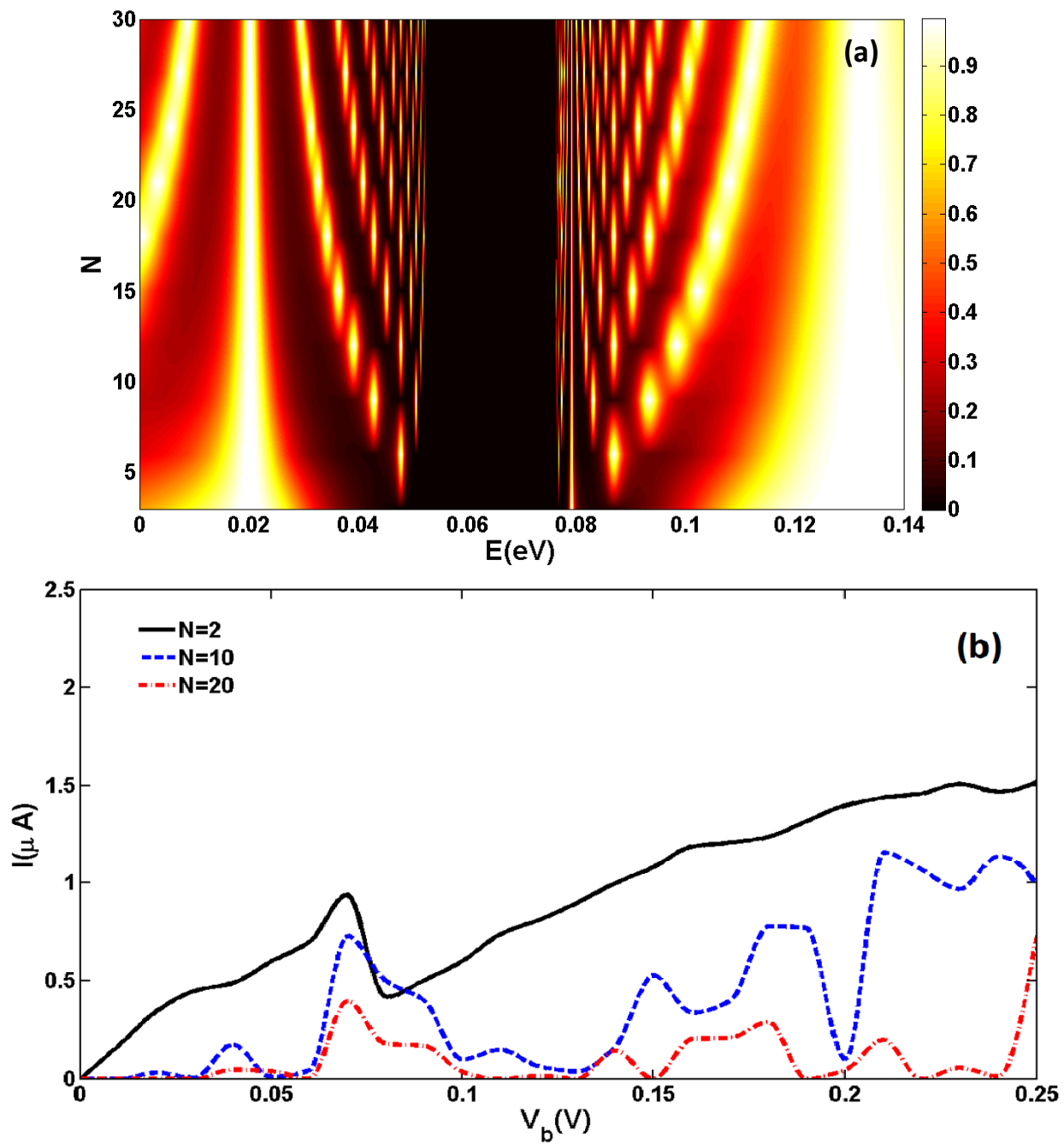

Fig. 4:

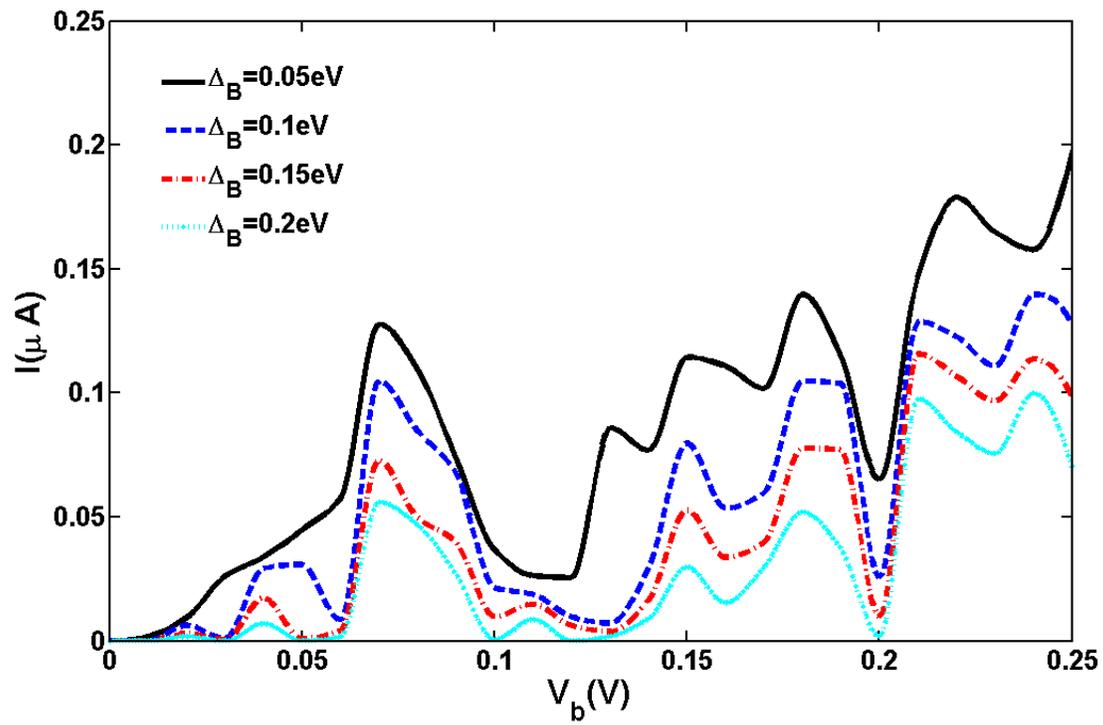

Fig. 5:

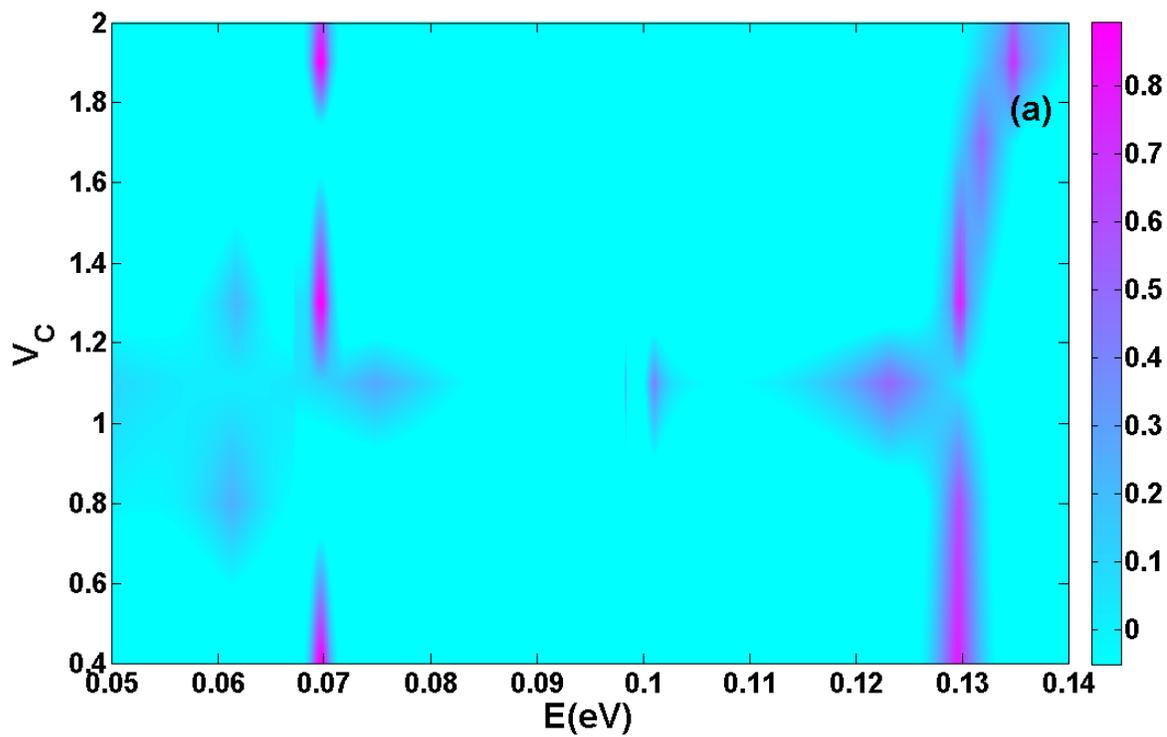

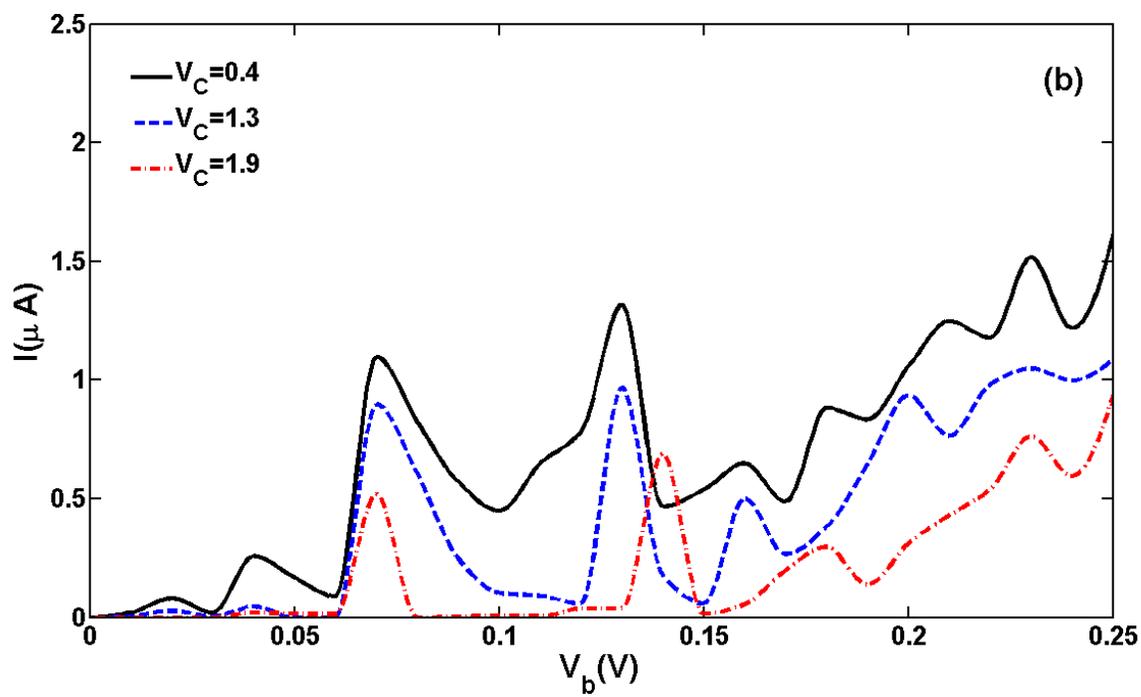

Fig. 6:

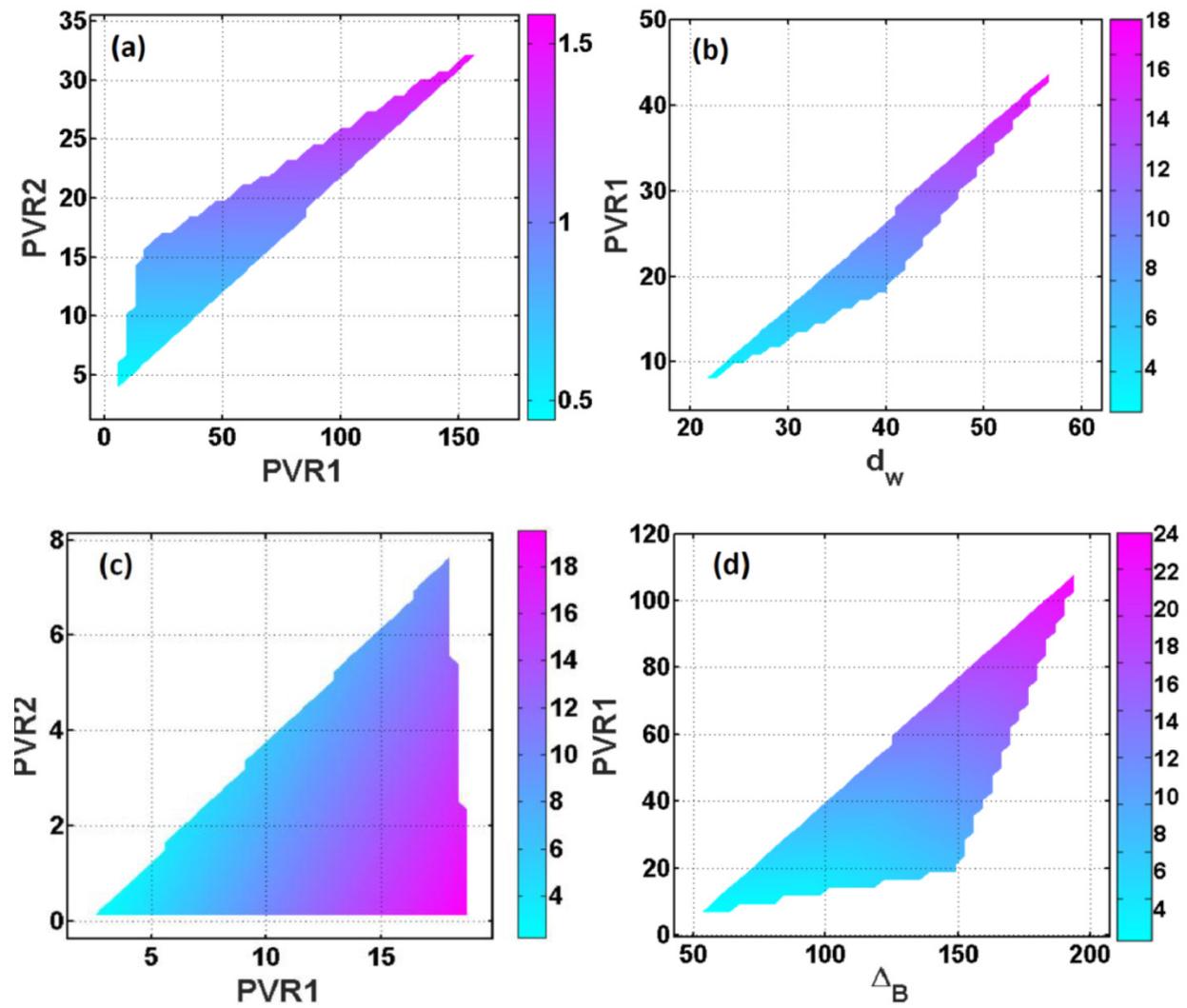